# PSEUDOSCOPIC IMAGING BY MEANS OF A HOLOGRAPHIC SCREEN

José J. Lunazzi
Campinas State University, Physics Institute
C.P. 6165 , 13084-100 Campinas-SP-Brazil
bitnet: lunazzi at bruc (lately lunazzi -at- ifi.unicamp.br)

## ABSTRACT

Pseudoscopic enlarged images are obtained by projecting diffraction-encoded images onto a diffractive screen

## INTRODUCTION

Pseudoscopic images can be produced by using a process of color-encoding/decoding of views generated by common diffraction gratings[1,2]. The necessary elements are: a common white object illuminated under white light, a diffraction grating for the encoding of views at the first diffraction order, and a lens for projecting the encoded image on a second grating. By keeping all the elements and distances symmetrical to the optical center of the lens, a pseudoscopic image is obtained as one of the diffracting beams. The whole process can be easily explained in terms or point symmetry. Quality in the image should be perfect due to the symmetric process, but is affected by aberration effects of the lens and from deformation and scattering at the gratings. Clear images can be seen in front of the second grating, in a different wavelength from each different point of view. If the object has moving parts going back and forth, the effect of being watching a fantastic world is remarkable, since it makes the observer to loose the reference of what is in front and what is in the back.

## 2. PROCEDURE FOR OBTAINING ENLARGED IMAGES

Almost the same properties can be obtained from an enlarged image if a holographic screen[3,4,5] is used in substitution to the second grating. The enlarging limit is just determined by the available brightness. The result is very impressive and looks as if a Benton hologram was made, just different in that the typical wavelength seguence goes from the vertical displacements case to the case of horizontal displacements. Although more analytical calculations are necessary to demonstrate that the resulting image keeps all the pseudoscopic proportionality after the enlarging process, their appearence allows us to presume for that.

## 3. EXPERIMENTAL RESULTS

We show at Figure 1 the photograph of the left and right point of view from an image that was enlarged by a factor of five, using a holographic screen made on 30x40 cm2 AGFA 8E75 film whose reference angle was 45 degrees. Although this picture appears as a stereo pair, the continuous field of view obtained was of 20cm by looking at a distance of 85cm. Distance from the projecting system to the screen was 1,2m and the image appeared at 9cm in front of the screen. The object was the luminous filament of a domestic lamp, about 2,7cm in their horizontal extension and of polygonal shape.

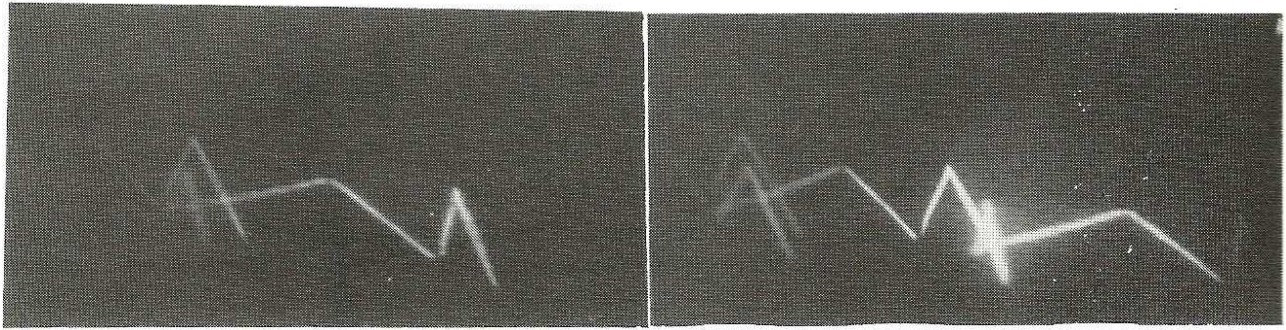

*Fig.1: Stereo photograph of the pseudoscopic image of the filament of a lamp. A second-order diffracted image appears at the left side of both views, while a zero order image may be seen at the right side of right view only.*

## 4. ACKNOWLEDGEMENTS

We acknowledge the Brazilian National Council for Research-CNPq for fellowships that helped the author and the student Jorge H. F. Puente to work on this subject, and to the Found of Assistance to Teaching and Research-FAEP-UNICAMP for its founding.